# Design and fabrication of an intrinsically gain flattened Erbium doped fiber amplifier

B. Nagaraju[1], M. C. Paul[2], M. Pal[2], A. Pal[2], R. K. Varshney[1], B. P. Pal[1, *], S. K. Bhadra[2], G. Monnom[3], and B. Dussardier[3]

[1]Physics Department, Indian Institute of Technology Delhi, New Delhi 110016, INDIA
[2]Central Glass and Ceramic Research Institute, Jadavpore, Kolkata 700032, INDIA
[3] Laboratoire de Physique de la Matière Condensée, Université de Nice Sophia Antipolis, Centre National de la Recherche Scientifique, Parc Valrose, F 06108 Nice CEDEX 2, FRANCE

**Abstract:**
We report design and subsequent fabrication of an intrinsically gain flattened Erbium-doped fiber amplifier (EDFA) based on a highly asymmetrical and concentric dual-core fiber, inner core of which was only partially doped. Phase-resonant optical coupling between the two cores was so tailored through optimization of its refractive index profile parameters that the longer wavelengths within the C-band experience relatively higher amplification compared to the shorter wavelengths thereby reducing the difference in the well-known tilt in the gains between the shorter and longer wavelength regions. The fabricated EDFA exhibited a median gain $\geq 28$ dB (gain excursion below $\pm 2.2$ dB within the C-band) when 16 simultaneous standard signal channels were launched by keeping the I/P level for each at $-20$ dBm/channel. Such EDFAs should be attractive for deployment in metro networks, where economics is a premium, because it would cut down the cost on gain flattening filter head.

* Corresponding author. Tel.: +91-11-26591327
Fax: +91-11- 2658 1114
*E-mail address*: bppal@physics.iitd.ernet.in

# 1. Introduction:

Erbium doped fiber amplifiers (EDFAs) find extensive applications in present optical communication systems because of their high gain, low noise and high-speed response. EDFAs also exhibit large gain bandwidth and a single EDFA can amplify large amount of data without any gain narrowing effects. So, a single EDFA can be used to amplify several channels simultaneously in a dense wavelength division multiplexing (DWDM) system. However, the non-uniform gain spectrum in conjunction with the saturation effects of EDFAs cause increase in signal power levels and decrease in the optical signal-to-noise ratio (OSNR) to unacceptable values in systems consisting of cascaded chains of EDFAs [1]. These features could limit the usable bandwidth of EDFAs and hence the amount of data transmission by the system. Accordingly various schemes of gain equalizing filters (GEFs) such as Mach-Zehnder filter [2], acousto-optic filter [3], long-period fiber-grating [4], fiber-loop mirror [5, 6], side-polished fiber based filter [7] and so on have evolved in the literature. However, as is well known, one of the major drivers in a metro network design is low installation cost in addition to achieving low maintenance/ repair costs. Naturally one of the routes to achieve these objectives would be to use fewer components in the network. Use of an intrinsically gain flattened EDFA would cut down the cost on the GEF head. This motivated us to investigate design of a gain flattened EDFA by exploiting a wavelength filtering mechanism inherent in a co-axial dual-core fiber design scheme. There have been earlier reports in the literature of few schemes to achieve inherent gain flattening in an EDFA through a twin core EDF [8-10] and also a co-axial dual core fiber design [11]. However, these reported schemes are relatively complex; for example, the twin core fiber requires fabrication of two separate preforms followed by polishing and complex procedure to assemble them as a composite unit in a fiber draw tower, while the coaxial design [11] requires i) an additional component in the form of a mode converter and ii) doping outer core with erbium, which is more demanding for the well known MCVD process of fiber fabrication. In this paper, we present functional principle, design and fabrication details of a new coaxial dual-core EDF [12] that does not have aforementioned problems.

# 2. Theoretical Analysis:

A schematic diagram of the RIP of the proposed fiber design is shown in Fig. 1. It consists of two highly asymmetric cores, an inner core with small index contrast and a much thinner outer core with a large index contrast while a matched index cladding connects the two cores. The parameter $r_d$ represents doping radius of the inner core, which is the only core doped with Erbium. The fiber parameters $a$, $b$, $c$, $n_1$ and $n_2$ were optimized such that the fundamental modes corresponding to the isolated cores were phase-matched at a wavelength near about 1533 nm, which we refer to as the phase matching wavelength ($\lambda_P$) for resonant coupling between the fundamental modes of the two co-axial cores. Thus as the wavelength changes from below to above $\lambda_P$, the mode field profile of the composite structure would undergo a significant change. For signals centered at wavelengths much shorter than $\lambda_P$, a large fraction of the signal power resides in the outer core. Fractional power in the inner core increases with increase in wavelength and finally for wavelengths longer than $\lambda_P$, the fractional power in the inner core becomes more than that in the outer core. Since only the inner core would be doped with Erbium ions, signals at those wavelengths longer than $\lambda_P$ would significantly overlap with the erbium doped region, and hence experience relatively larger gain compared to wavelengths shorter than $\lambda_P$. As a result, the tilt in the gain spectrum between signals at the shorter and the longer wavelengths within the C-band would reduce leading to an effective flattening of the gain spectrum of the EDFA.

Figure 2 shows the wavelength variation of mode effective index ($n_{eff}$) for the core (isolated inner core) and the ring modes (isolated outer core), as well as those of the $LP_{01}$ and $LP_{02}$ modes of the composite coaxial fiber. It can be seen from the figure that $n_{eff}$ of the $LP_{01}$ mode is close to that of the ring mode at wavelengths shorter than $\lambda_P$, while for wavelengths longer than $\lambda_P$ it is closer to that of the core mode. Therefore, by optimizing the parameters for optimum $\lambda_P$ and doping erbium in the inner core, we can achieve an increased overlap of the modal field with the doped region for the longer signal wavelengths, thus enabling higher gain in that wavelength region. Spectral dependence of the fractional powers in the two cores is shown in Fig. 3.

In order to obtain the most suitable index profile parameters commensurate to ease in our targeted fabrication of an inherently gain flattened EDFA by the MCVD method, the gain and other important characteristics of a co-axial dual core EDF were modeled through the standard three-level rate equation model [13]. We have assumed forward pumping at 980 nm wavelength via the $LP_{02}$ mode and that the signal is launched into the $LP_{01}$ mode only. The model also included the wavelength dependent forward and backward traveling amplified spontaneous emission (ASE). ASE has been determined at 100 wavelength points, spaced 1 nm apart, in the wavelength range 1500-1600 nm and the propagation effects were calculated for each of these sample wavelengths. The coupled nonlinear differential equations, which govern the propagation of ASE, signal and pump powers along the length of the fiber, are given by [13]

$$\frac{dS_{ase}^{\pm}(\nu,z)}{dz} = \pm 2h\nu\gamma_e(\nu,z) \pm \left[\gamma_e(\nu,z) - \gamma_a(\nu,z)\right]S_{ase}^{\pm}(\nu,z) \tag{1}$$

$$\frac{dP_{p,s}(z)}{dz} = \left[\gamma_e(\nu_{p,s},z) - \gamma_a(\nu_{p,s},z)\right]P_{p,s}(z) \tag{2}$$

The emission and absorption factors, $\gamma_e$ and $\gamma_a$ are given by

$$\gamma_e(\nu,z) = \sigma_e(\nu,z)2\pi\int_0^{r_d} N_2(r,z)I(\nu,r)rdr \tag{3}$$

$$\gamma_a(\nu,z) = \sigma_a(\nu,z)2\pi\int_0^{r_d} N_1(r,z)I(\nu,r)rdr \tag{4}$$

where $r_d$ is the doping radius, $I(v, r)$ is the normalized intensity distribution at frequency $v$ and $\sigma_a$ and $\sigma_e$ are the wavelength-dependent absorption and emission cross-sections, respectively. Population densities in the ground ($N_1$) and excited ($N_2$) states respectively, are given by

$$N_2(r,z) = \rho_{Er} \frac{W_a(r,z)}{W_a(r,z) + W_e(r,z)} \tag{5}$$

$$N_1(r,z) = \rho_{Er} - N_2(r,z) \tag{6}$$

with $\rho_{Er}$ representing the concentration of $Er^{3+}$ ions while $W_a$ and $W_e$, respectively, are the total absorption and emission rates given by

$$W_a(r,z) = \left[ \frac{\sigma_a(v_s)}{hv_s} P_s(z) + \int_0^\infty \frac{\sigma_a(v)}{hv} S_{ase}(v,z)dv \right] I_s(r) + \left[ \frac{\sigma_a(v_p)}{hv_p} P_p(z) \right] I_p(r) \tag{7}$$

$$W_e(r,z) = \left[ \frac{\sigma_e(v_s)}{hv_s} P_s(z) + \int_0^\infty \frac{\sigma_e(v)}{hv} S_{ase}(v,z)dv \right] I_s(r) + A_{21} \tag{8}$$

where $A_{21}$ is the spontaneous emission rate $I_s$ and $I_p$ are the normalized intensities of the modes at the signal and pump wavelengths, respectively and $P_s$ and $P_p$ are the modal powers at the signal and pump wavelengths at a spatial location $z$ along the fiber; $S_{ASE}$ is the total of the forward and backward ASE power spectral densities. Modal intensities $I_{s,p}$ were obtained through the well-known Matrix method [14] and the variation of refractive index with wavelength was calculated using Sellemeir equations [15].

Equations (1 & 2) were solved over the bandwidth of 100 nm (1500-1600 nm) using fourth-order Runge-Kutta method with adaptive step size to obtain gain spectrum of the EDF. Equations (3 & 4) were solved by Simpson's method and considering the overlap integrals and the radial variation of the population density, transition rates and the modal field profiles at 41 points within the doping radius. Since both forward and backward ASEs were considered, the effect of the backward traveling ASE on the population inversion influences the propagation of the forward propagating light and vice versa, hence a large number of iterations were required to obtain a stable solution.

### 3. Fiber Fabrication and Results:
Using above-mentioned approach, optimized parameters of the proposed fiber design were calculated. Keeping in view fabrication constraints of the MCVD method, one such set (with respect to the RIP of Fig. 1) was $r_d$, $a$, $b$, $c$, $d$ as 1.5, 5.25, 13, 14.8, and 62.5 all in μm, respectively; refractive indices $n_0$, $n_1$, $n_2$ respectively, were 1.44402, 1.45327, and 1.4617 at $\lambda$ = 1550 nm; $Er^{+3}$-concentration was chosen to be $1.75\times10^{25}$ ions/$m^3$. Based on this design as a target, fabrication recipe was defined and an EDF was fabricated using conventional MCVD technique followed by solution doping. From the fabrication point of view, the critical steps involved were i) to achieve the thickness of a thin ring core of width ~ 1.40 µm, which required appropriate adjustment of the flow of $SiCl_4$ with respect to refractive index modifiers like $GeCl_4$ through suitable number of deposition passes as well as control of the burner traversal speed, ii) tight temperature control during deposition of the inner core layers for retaining certain thin soot layers of $SiO_2$-$GeO_2$-$Al_2O_3$ un-sintered for partial doping (8 ~ 9%) of the inner core later with $Er^{3+}$, and iii) to maintain the refractive index same throughout the core with or without the $Er^{3+}$ ions.

Coupling of the maximum fractional power to $LP_{01}$ mode at signal wavelengths is an important issue in such a dual-core fiber. The excitation efficiency between standard transmission fibers and the $LP_{01}$ and $LP_{02}$ modes of the EDF would vary significantly with wavelength within the C-band due to large variation in the modal field profiles. However, the scenario is completely different when we taper the EDF and then splice it to the transmission fiber. In Fig. 4, we show the excitation efficiencies of $LP_{01}$ and $LP_{02}$ modes of the EDF across C-band for two different down-tapering levels of 5% and 10%. This figure clearly indicates that by slightly tapering the EDF, the coupling efficiency to the $LP_{01}$ mode can be maximized (> 95% power coupled to $LP_{01}$ mode for 10% taper). We should also mention that at the pump wavelength of 980 nm, fiber would support several modes. Since the pump wavelength is much smaller than $\lambda_p$, most of the fractional power of the $LP_{01}$ mode resides in the outer ring and hence the coupling efficiency of standard transmission fiber to the $LP_{01}$ mode would be small due to poor mode overlap. Since our EDF is doped in the inner core region, for maximum pump efficiency, it would be desirable to couple power to the $LP_{02}$ and $LP_{03}$ modes. Fortuitously, the coupling from standard transmission fiber to $LP_{02}$ and $LP_{03}$ modes is large due to large overlap of modal profiles, which ensures efficient pumping at the pump wavelength and no intermediate mode conversion device is required. For example, if the transmission fiber is the standard SMF-28, then about 95% power would couple to the $LP_{02}$ mode, 4% to the $LP_{03}$ and less than 0.5% to the $LP_{01}$ mode. The remaining power is distributed among the other asymmetric modes. Moreover, this coupling efficiency would be further enhanced when the EDF is tapered and spliced to SMF-28, e.g. for a 10% tapering, the coupling efficiency to $LP_{02}$ mode increases to 98% from ~ 95% for the untapered case. Thus ~ 10% tapering of the EDF will be useful for achieving significant excitation efficiency between a standard G652 type transmission fiber and the desired modes of the EDF.

An optical micrograph of the fabricated preform is shown in Fig. 5(a) and the corresponding measured RIP is shown in Fig. 5(b). The RIP was measured using a fiber analyzer. The so realized RIP was close to the designed one except for small profile perturbations typical in fibers fabricated by the MCVD process. Figure 6 shows the measured gain and noise figure as a function of wavelength for the fabricated coaxial fiber; some improvements in the noise figure at longer wavelengths could be seen due to the increased overlap between the pump and the signal modes at longer wavelengths. Gain variation across the C-band was found to be more than the designed one, which is attributable to small variations in the fabricated fiber RIP parameters from the one that was designed. A very precise comparison is, in any case, difficult due to lack of sufficient precision inherent in measurement instruments for estimating various parameters of the fiber RIP and the dopant level. Nevertheless, initial design parameters could be used as a very useful guideline to define fabrication recipe. Further work is in progress to achieve still better degree of gain flattening and for realization of an inherently gain-flattened EDFA in the L-band also.

## 4. Conclusion:

We have proposed functional principle, design and fabrication of an inherently gain flattened EDFA through exploitation of the resonant coupling of modes between the two highly asymmetric cores of a co-axial dual core fiber, whose only inner core was doped partially with $Er^{3+}$ ions. The RIP parameters of the fiber were optimized at the design stage to minimize gain excursion within the C-band, which formed target specifications for subsequent fabrication of the EDF by the MCVD method. Median gains $\geq$ 28 dB with a noise figure of 4 ~ 6 dB were demonstrated under multi-signal channel operation with 16 signal channels within the C-band. Such intrinsically gain flattened EDFAs should be attractive for reduced cost-driven transparent metro networks since it would imply a cost cutting on the GEF head of EDFAs required for such networks. We believe that with further optimization and perfection of the fiber fabrication process, the gain excursion could be brought down to a still lower figure.

**Acknowledgement:**

Authors acknowledge partial support of the work through the P2R Indo-French Networking Project funded by the Department of Science and Technology (DST), Govt. of India and the French Ministry of research. H. S. Maiti, Director of CGCRI, Kolkata is thanked for his keen interest and support.

**References**
1. A. Srivastava and Y. Sun, Erbium doped fiber amplifiers for dynamic optical networks, in: B.P. Pal (Ed.) Guided wave optical components and devices: basics, technology and applications, Chapter 12, Burlington: Academic Press, Elsevier, 2006.
2. J. Y. Pan, M. A. Ali, A. F. Elrefaie, and R. E.Wagner, "Multiwavelength fiber-amplifier cascades with equalization employing Mach-Zehnder optical filter", *IEEE Photon. Technol. Lett.,* **7** (1995), 1501-1503.
3. H.S. Kim, S.H. Yun, H. K. Kim, N. Park, and B.Y. Kim, "Actively Gain-flattened erbium-doped fiber amplifier over 35 nm by using all-fiber acousto-optic tunable filters", *IEEE Photon. Technol. Lett.,* 10 (1998), 790-792.
4. A. M. Vengsarkar, P. J. Lemaire, J. B. Judkins, V. Bhatia, T. Erdogan, and J. E. Sipe, "Long-period fiber gratings as band-rejection filters", *IEEE J. Lightwave Technol.,*14 (1996), 58-64.
5. S. Li, K. S. Chiang, and W. A. Gambling, "Gain flattening of an erbium doped fiber amplifier using a high-birefringence fiber loop mirror", IEEE Photon. Technol. Lett. 13 (2001), 942-944.
6. N. Kumar, M.R. Shenoy, and B.P. Pal, "A standard fiber-based loop mirror as a gain-flattening filter for erbium-doped fiber amplifiers", IEEE Photon. Technol. Lett. 17 (2005), 2056-2058.
7. R. K. Varshney and B. Nagaraju, A. Singh, B. P. Pal, and A. K. Kar, "Design and realization of an all-fiber broadband tunable gain equalization filter for DWDM signals", *Optics Express,* 15 (2007), 13519-13530.
8. R. I. Laming, J. D. Minelly, L. Dong, and M. N. Zervas, "Twin-core erbium doped fiber amplifier with passive spectral gain equalization", *Electron. Lett.* 29 (1993), 509-510.
9. B. Wu, and P. L. Chu, "A twin-core erbium-doped fiber amplifier", *Opt. Commun.,* 110 (1994), 545-548.
10. Y. B. Lu, and P. L. Chu, "Gain flattening by using dual-core fiber in erbium-doped fiber amplifier", *IEEE Photon. Technol. Lett.,* 12 (2000), 1616-1617.
11. K. Thyagarajan, and J. Kaur, "A novel design of an intrinsically gain flattened erbium doped fiber", *Opt. Commun.,* 183 (2000), 407-413.
12. B. Nagaraju, M. C. Paul, M. Pal, A. Pal, R. K. Varshney, B. P. Pal, S. K. Bhadra, G. Monnom, and B. Dussardier, "Design and realization of an inherently gain flattened Erbium doped fiber amplifier" Paper JTuA86, CLEO, San Jose, May 4-9, 2008.
13. B. Pedersen, "Small-signal erbium-doped fiber amplifiers pumped at 980 nm: a design study", *Optical and Quantum Electronics,* 26 (1994), S273-S284.
14. K. Thyagarajan, S. Diggavi, A. Taneja, and A. K. Ghatak, "Simple numerical technique for the analysis of cylindrical symmetric refractive index profile optical fibers", *Appl. Opt.,* 30 (1991), 3877-3879.
15. M. J. Adams, An Introduction to Optical Waveguides*,* Chichester: Wiley, 1981.

**Figure Captions:**

Fig.1. Schematic of the refractive index profile (RIP) of the proposed fiber.

Fig.2. Variation of mode effective indices of core mode, ring mode, and the $LP_{01}$ and $LP_{02}$ modes of the fiber.

Fig.3. Wavelength dependence of the fractional power within the two individual cores.

Fig.4. Variation of excitation efficiencies of the $LP_{01}$ and $LP_{02}$ modes across the C-band with tapering and fusion splicing the EDF with SMF-28 fiber.

Fig.5. (a) Microscopic view of the cross section of dual core fiber preform. (b) Measured refractive index profile of the fabricated inherently gain flattened EDF.

Fig.6. The measured gain and noise figure of the fabricated EDF for the sample length of 12 meter, which was the optimum.

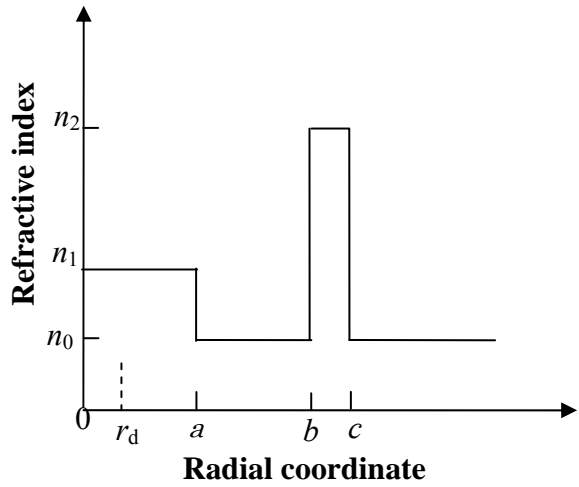

**Fig.1**

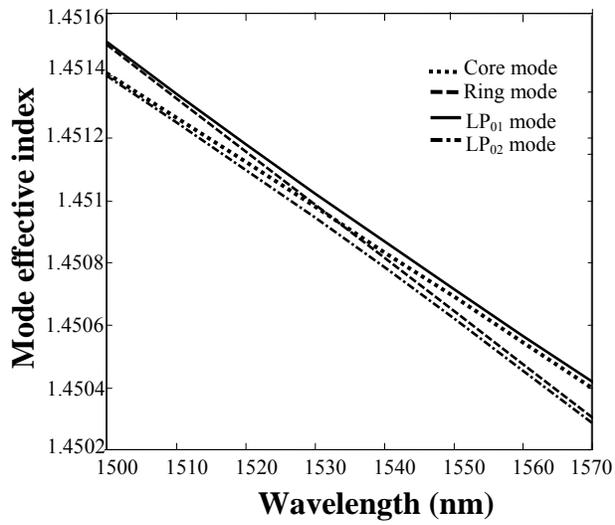

**Fig.2**

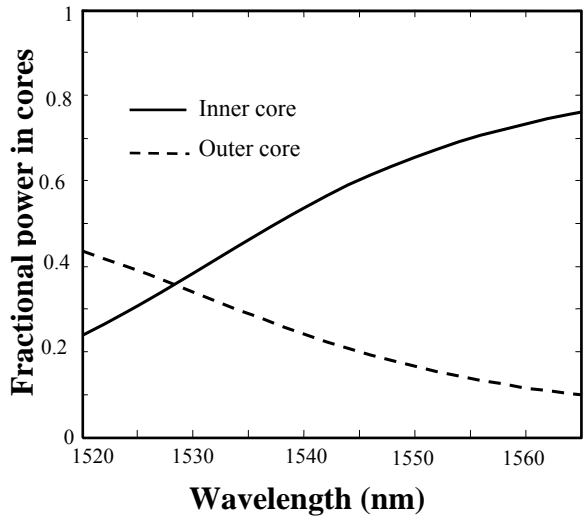

**Fig. 3**

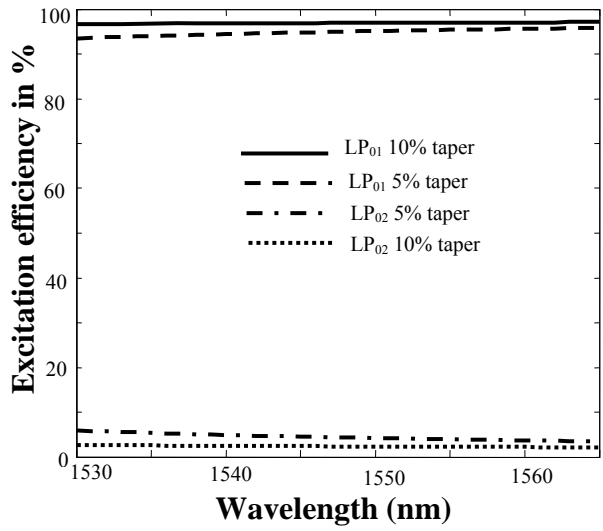

**Fig. 4**

(a)

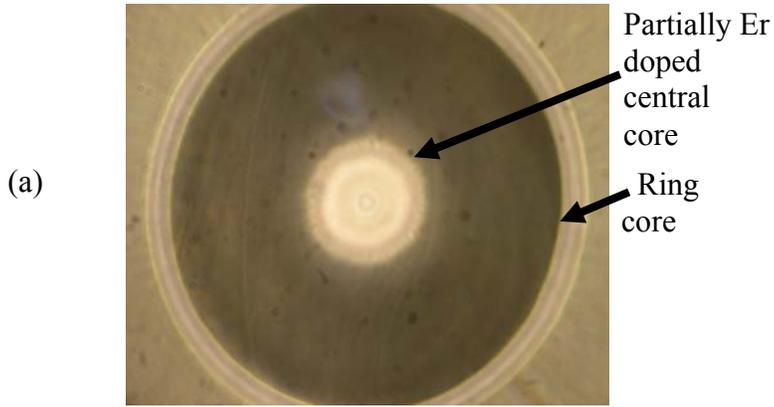

Partially Er doped central core

Ring core

(b)

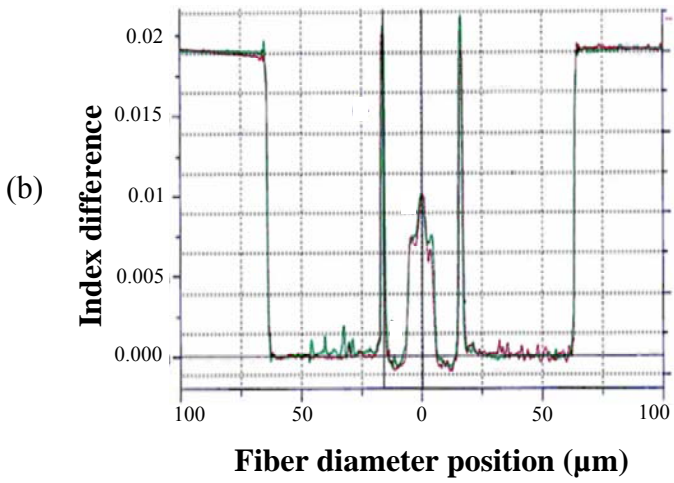

**Fig. 5**

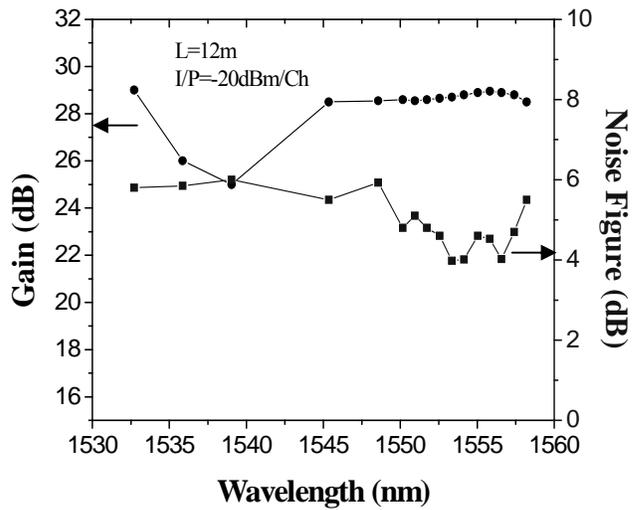

**Fig. 6**